\documentclass[12pt]{article}
\usepackage{pslatex}
\usepackage{graphics}
\begin{document}

\title{The Production Function}
\author{Guido Fioretti \thanks{Contact address: Gui\-do Fio\-ret\-ti,
  via di Cor\-ti\-cel\-la 23, I - 40128 Bo\-lo\-gna. E-mail: fio\-ret\-ti@cs.unibo.it.} \\ University of Bo\-lo\-gna \\ Department of Business Economics \\ Department of Computer Science}
\maketitle

\begin{abstract}
Productions functions map the inputs of a firm or a productive system onto its outputs. This article expounds generalizations of the production function that include state variables, organizational structures and increasing returns to scale. These extensions are needed in order to explain the regularities of the empirical distributions of some economic variables.
\end{abstract}


\section{Introduction}

In the course of its development, the economic theory of production explored paths that are at odds with general equilibrium theory. Some of these research directions may be crucial in order to explain the regularities that have been discovered regarding several economic variables, notably the distribution of the size of firms. In particular, this article expounds some extensions of the  function that maps the production factors onto the products of a firm, the \emph{production function}.

Most of the issues expounded in the rest of this paper are little known among economists themselves. If they are, they are known in a lengthy, often verbal form. By compacting them in mathematical language, this article hopes to be helpful to those physicists wishing to apply their methods to  economic problems.

The rest of this article is organized as follows. Section~\ref{sec:iso} adds state variables to the conventional form of the production function. Section~\ref{sec:organization} explores the possibility of production functions that take account of the structural relations within a firm or an economic system. Section~\ref{sec:increasing} explores the possibility of generalized increasing returns to scale with reference to the empirical regularities that have been observed in the distribution of the size of firms. Finally, section~\ref{sec:context} frames the previous issues with respect to the development of economics and its relationships with other disciplines.

\section{State variables in the production function} \label{sec:iso}

Any science is tempted by the naive attitude of describing its object of enquiry by means of input-output representations, regardless of state. For instance, in the first half of the XIX century it was fashionable for psychologists to study the behavior of animals in terms of stimuli such as hunger or pain (the input) and the response in terms of unrest or aggressiveness (the output). No attention was paied to what was going on inside the brain (the state).

Tipically, microeconomics describes the behavior of firms by means of a \emph{production function}:

\begin{equation}
\mathbf{y} = f(\mathbf{x})  \label{eq:basic}
\end{equation}
where $\mathbf{x} \in \mathcal{R}$ is a $p \times 1$ vector of production factors (the input) and  $\mathbf{y} \in \mathcal{R}$ is a $q \times 1$ vector of products (the output).

Both $\mathbf{y}$ and $\mathbf{x}$ are flows expressed in terms of  physical magnitudes per unit time. Thus, they may refer to both goods and services.

Clearly, eq.~\ref{eq:basic} is independent of state. Economics knows state variables as \emph{capital}, which may take the form of financial capital (the financial assets owned by a firm), physical capital (the machinery owned by a firm) and human capital (the skills of its employees). These variables should appear as arguments in eq.~\ref{eq:basic}.

This is done in the \emph{Georgescu-Roegen production function} \cite{georgescuroegen70}, which may be expressed as follows:

\begin{equation}
\mathbf{y} = f(\mathbf{k},\mathbf{x})  \label{eq:georgescu-roegen}
\end{equation}
where $\mathbf{k} \in \mathcal{R}$ is a $m \times 1$ vector of capital endowments, measured in physical magnitudes. Without loss of generality we may assume that the first $m_p$ elements represent physical capital, the subsequent $m_h$ elements represent human capital and the last $m_f$ elements represent financial capital, wuth $m_p + m_h + m_f = m$.

Contrary to input and output flows, capital is a stock. Physical capital is measured by physical magnitudes such as the number of machines of a given type. Human capital is generally proxied by educational degrees. Financial capital is measured in monetary terms.

Georgescu-Roegen called the stocks of capital  \emph{funds}, to be contrasted to the \emph{flows} of products and production factors. Thus, Georgescu-Roegen's production function is also known as the \emph{flows-funds model}.

Georgescu-Roegen's production function is little known and seldom used, but macroeconomics often employs aggregate production functions of the following form:

\begin{equation}
Y=f(K,L)  \label{eq:macro}
\end{equation}
where $Y \in \mathcal{R}$ is aggregate income, $K \in \mathcal{R}$ is aggregate capital and $L \in \mathcal{R}$ is aggregate labor. Though this connection is never made, eq.~\ref{eq:macro} is a special    case of eq.~\ref{eq:georgescu-roegen}.

The examination of eq.~\ref{eq:macro} highlighted a fundamental difficulty. In fact, general equilibrium theory requires that  the remunerations of production factors are proportional to the corresponding partial derivatives of the production function. In particular, the wage must be proportional to $\partial f / \partial L$ and the interest rate must be proportional to $\partial f / \partial K$. These partial derivatives are uniquely determined if $\mathrm{d}f$ is an exact differential.

If the production function is eq.~\ref{eq:basic}, this translates into requiring that:

\begin{equation}
\frac{\partial ^{2}f}{\partial x_i \partial x_j} =  \frac{\partial ^{2}f}{\partial x_i \partial x_j} \;\;\; \forall i,j    \label{eq:conditionx}
\end{equation}
which are surely satisfied because all $x_i$ are flows so they can be easily reverted.

If the production function is expressed by eq.~\ref{eq:georgescu-roegen} but $m=1$ the following conditions must be added to conditions~\ref{eq:conditionx}:

\begin{equation}
\frac{\partial ^{2}f}{\partial k \partial x_i} =  \frac{\partial ^{2}f}{\partial x_i \partial k} \;\;\; \forall i    \label{eq:conditionxk}
\end{equation}

Conditions~\ref{eq:conditionxk} are still surely satisfied because there is only one capital good. However, if $m>1$ the following conditions must be added to conditions~\ref{eq:conditionx}:

\begin{equation}
\frac{\partial ^{2}f}{\partial k_i \partial x_j} =  \frac{\partial ^{2}f}{\partial x_j \partial k_i} \;\;\; \forall i,j   \label{eq:conditionxkk}
\end{equation}

\begin{equation}
\frac{\partial ^{2}f}{\partial k_i \partial k_j} =  \frac{\partial ^{2}f}{\partial k_j \partial k_i} \;\;\; \forall i,j    \label{eq:conditionk}
\end{equation}

Conditions~\ref{eq:conditionxkk} and~\ref{eq:conditionk} are \emph{not} necessarily satisfied because each derivative depends on all stocks of capital $k_i$. In particular, conditions~\ref{eq:conditionxkk} and~\ref{eq:conditionk} do not hold if, after capital $k_i$ has been accumulated in order to use the technique $i$, capital $k_j$ is accumulated in order to use the technique $j$ but, subsequently, production reverts to technique $i$. This possibility, known as \emph{reswitching of techniques}, undermines the validity of general equilibrium theory \cite{robinson53-54} \cite{sraffa60} \cite{garegnani70}.

For many years, the reswitching of techniques has been regarded as a theoretical \textit{curiosum}. However, the recent upsurge of biodynamic agriculture or the resurgence of coal as a source of energy may be regarded as instances of reswitching.

Finally, it should be noted that as any input-state-output representation,  eq.~\ref{eq:georgescu-roegen} must be complemented by the dynamics of the state variables:

\begin{equation}
\dot{\mathbf{k}} = g(\mathbf{k},\mathbf{x},\mathbf{y})  \label{eq:k}
\end{equation}
which updates the vector $\mathbf{k}$ in eq.~\ref{eq:georgescu-roegen} making it dependent on time.

In the case of the aggregate production function~\ref{eq:macro}, eq.~\ref{eq:k} combines with~\ref{eq:macro} to constitute a \emph{growth model}. In the case of the microeconomic production function~\ref{eq:georgescu-roegen}, explicitating eq.~\ref{eq:k} requires investigations of the strategies pursued by a firm.

\section{The production function of an organization} \label{sec:organization}

A production function in the form of eq.~\ref{eq:georgescu-roegen} does not distinguish elementary units within an organization. This may be a serious shortcoming if the structure of interactions between elementary units affects the final outcome. Unstructured production functions have the same limitations as using compact chemical formulae  to represent allotropes.

Let us consider an organization composed by $n$ units. Let $\mathbf{K}$ denote a $m \times n$ matrix of capital endowments. The  $i$-th column of $\mathbf{K}$,  denoted by $\mathbf{k}^i$, represents the $m$ capital endowments of unit $i$.

Let $\mathbf{y} = F(\mathbf{K}, \mathbf{x})$ denote the production function of the organization, where $\mathbf{x}$ is a $p \times 1$ vector and $\mathbf{y}$ is a $q \times 1$ vector. The organizational production function $F$ arises from the interaction of $n$  functions $\mathbf{y}^i = f^{i}(\mathbf{k}^i, \mathbf{x}^i, \mathbf{y}^j \ldots)$ with $j \in \{1, \ldots n\}$. The $p_i \times 1$ vector  $\mathbf{x}^i$ entails the inputs that reach unit $i$. The  $q_i \times 1$ vector  $\mathbf{y}^i$ entails the outputs of unit $i$.

Vector $\mathbf{x}$ can be seen as arising from concatenation of the $n$ vectors $\mathbf{x}^i$. Likewise, $\mathbf{x}$ can be seen as arising from concatenation of the $n$ vectors $\mathbf{x}^i$. With this convention, $\sum_{i=1}^{n} p_i = p$ and $\sum_{i=1}^{n} q_i = q$. However, in general the inputs to the organization do not reach all units. Likewise, not all the outputs of all units are necessarily represented in the organizational output. Thus, some components of $\mathbf{x}$ and some components of $\mathbf{y}$ may be zero.

Since organizational units may feed their outputs into other units, all outputs $\mathbf{y}^1 \ldots \mathbf{y}^n$, including  $\mathbf{y}^l$, can be arguments of $f^l$. Which particular $\mathbf{y}^j$s appear in each $f^{i}$ depends on the structure of the organization.

Organizational structures are often quite intricated. Thus, only in a few cases it is possible to derive a closed form of $F$ from knowledge of the $f^i$s  and  their connections.  In general, local linearized descriptions are a more affordable goal.

Let us define a  $q \times p$  matrix $\Gamma(\overline{\mathbf{x}})$ entailing the derivatives of the outputs of all units with respect to the  inputs evaluated at  $\overline{\mathbf{x}}$.  If  input $i$ does not reach  unit $j$, let us stipulate that $\gamma_{ij} = 0$. Furthermore, let us define a $q \times q$ matrix $\Omega(\overline{\mathbf{y}})$ entailing the derivatives of the outputs of all units with respect to the outputs of units that feed into them evaluated at $\overline{\mathbf{y}}$. If the output $i$ does not feed back into unit $j$, let us stipulate that $\omega_{ij} = 0$.

With these matrices, the differentials of inputs and outputs are linked by $d \mathbf{y} = \Gamma(\overline{\mathbf{x}}) \, d\mathbf{x} + \Omega(\overline{\mathbf{y}}) \, d\mathbf{y}$. Thus, the organizational production function linearized in a neighborhood of $(\overline{\mathbf{x}}, \overline{\mathbf{y}})$ takes the form:

\begin{equation} \label{eq:local}
\overline{F}(\mathbf{x}) = \frac{\Gamma(\overline{\mathbf{x}})}{\mathbf{I} - \Omega(\overline{\mathbf{y}})} \: \mathbf{x}
\end{equation}

Matrix $\Omega$ entails the structure of connections between organizational units. If these connections are such that they produce cycles, particular sequences of operations can be repeated indefinetly giving rise to a \emph{routine} \cite{nelson-winter82}. Note that according to this definition routines belong to the organization but not to their members, who may even be unaware of them. Thus, this formalization may capture fuzzy but fundamental concepts such as ``organizational culture'' or, at a more aggregate level, ``social capital''.

In the parlance of artificial intelligence,  the circuits embedded in $\Omega$ implement a \emph{distributed memory} that does not reside in any single unit but belongs to the organization as a whole. No single unit is able to retrieve this knowledge, but the organization as a whole does if certain environmental stimuli trigger particular routines. On the contrary, capital endowments $\mathbf{K}$ can be seen as \emph{localized memories} owned by the single units. Social science prefers the terms \emph{distributed knowledge} and \emph{localized knowledge}, respectively.

Thus, the organizational production function has a lot in common with connectionist models. In particular, let us make the following assumptions:

\begin{enumerate}

\item Each component of the input vector  $\mathbf{x}$ is distributed to one and only one unit. Thus, $p_i = 1 \: \forall i$ and  $p=n$. \label{ass:input}

\item Each unit produces only one good $y^i$, which is a component of the output vector $\mathbf{y}$. Thus, $q_i = 1 \: \forall i$ and $q=n$. \label{ass:output}

\item The production functions of organizational units take the form $y^i = x^i + \sum_{j=1}^{n} k_ij y^j$, where the weights $k_ij$ are the elements of the  $n \times n$ matrix of capital endowments $\mathbf{K}$. \label{ass:function}

\end{enumerate}

With assumptions~\ref{ass:input}, \ref{ass:output} and~\ref{ass:function}, $\Gamma \equiv \mathbf{I}$ and $\Omega \equiv \mathbf{K}$. Thus, eq.~\ref{eq:local} takes the form $\mathbf{I} / \mathbf{I} - \mathbf{K}$.
This equation characterizes the short-term response of associative memories implemented on neural networks \cite{kohonen89}.

Similarly to eq.~\ref{eq:k}, there must be a mechanism for updating the capital endowments of organizational units:

\begin{equation}
\dot{\mathbf{k}}^i = g^{i}(\mathbf{K},\mathbf{x},\mathbf{y}) \;\; \forall i  \label{eq:ki}
\end{equation}

Eq.~\ref{eq:ki} is quite general. In fact, it says that the capital endowments of unit $i$ may change depending on its own capital endowments as well as the capital endowments of other units, depending on all inputs to the organization and depending on all outputs of the organization.

If assumptions~\ref{ass:input}, \ref{ass:output} and~\ref{ass:function} hold, the analogy between organizational production functions and neural networks can be further pursued.  In fact, eq.~\ref{eq:ki} corresponds to the rule for updating the weights of the neurons. In particular, both the back-propagation algorithm employed in supervised networks and the rules employed in unsupervised networks are such that  $\dot{\mathbf{k}}^i = \alpha \, \mathbf{x} \, y_i$, with $\alpha \in \Re$.

Indeed,  hierarchies are often managed  like supervised neural networks. In fact, resources are assigned to subordinated depending on the distance between current output and a target fixed by the supervisor. On the contrary, regional or national economies are managed by decentralized decisions depending on local performance, similarly to unsupervised neural networks.

Fioretti exploited the above similarities in order to model the recognition of the potentialities of novel technologies by firms and productive systems \cite{fioretti04} \cite{fioretti06}, but this approach is new to economics. In general, economics ignores organizational production functions except in the limited case of positive externalities between firms.  In fact, economics resorts to positive externalities in order to explain geographical agglomeration of industries \cite{krugman91} \cite{fujita-krugman-venables99} or, more in general, economic growth in spite of non-increasing returns to scale of single firms \cite{romer87} \cite{romer90}.

In these applications, the  units are firms and the organizational production function is the macroeconomic production function. Since firms are independent from one another, the organizational production function is such  that $\Gamma \equiv \mathbf{I}$.

Positive externalities are supposed to emanate from firms because firms provide workers with human capital that they diffuse when they move to other firms. Through a mean field, firms feed back into other firms. In regional models, each firm feeds back into firms that are geographically close. In growth models, each firm feeds back into all other firms. Thus, in regional models the elements of $\Omega$ depend both on the relevance of human capital and the geographical distance between any two firms. On the contrary, in growth models the elements of $\Omega$ depend only on the relevance of human capital.

A further simplification is generally made by assuming that there exists only one kind of human capital. With this assumption, in regional models the rows of $\Omega$ are proportional to one another by a factor depending on physical distance. On the contrary, in growth models the rows of $\Omega$ are all equal to one another.

Thus, growth models are generally aggregate models. The positive feedbacks are captured by a term increasing with time that is multiplied to a standard \emph{Cobb-Douglas} production function:

\begin{equation}
Y = A(t)\,K^\alpha \,L^{1-\alpha}    \label{eq:cobb-douglas}
\end{equation}
where $\alpha \in \Re$. As in eq.~\ref{eq:macro}, $Y$ is aggregate output, $K$ is aggregate capital and $L$ is aggregate labor. The term $A(t)$ is such that $\dot{A} >0$.

This function is compatible with general equilibrium theory because it has constant returns to scale. However, the term $A(t)$ is able to account for economic growth, allegedly arising out positive externalities between firms operating with decreasing returns to scale.

Macroeconomic data apparently fit into eq.~\ref{eq:cobb-douglas}. However, it can be shown that eq.~\ref{eq:cobb-douglas} may fit reality because of algebraic reasons, not because of economic reasons \cite{hogan58} \cite{shaikh74} \cite{moss}.

By definition, at any time aggregate income $Y$ is equal to the sum of aggregate wages $W$ and aggregate profits $\Pi$:

\begin{equation} \label{eq:YWP}
Y(t) \equiv W(t) + \Pi(t)
\end{equation}

This identity can be divided by the amount of labor $L(t)$ in order to be expressed as $y(t) \equiv w(t) + r(t)k(t)$, where $y(t) = Y(t)/L(t)$ is the income/labor ratio, $w(t)=W(t)/L(t)$ is the wage rate, $r(t)=\Pi(t)/K(t)$ is the profit rate and $k(t)=K(t)/L(t)$ is the capital/labor ratio. By differentiating one obtains $\dot{y}/y \equiv w/y(\dot{w}/w) + rk/y(\dot{r}/r) + rk/y (\dot{k}/k)$. By denoting the share of profits $\alpha=rk/y$ and the share of wages $1-\alpha=w/y$, eq.~\ref{eq:YWP} can be expressed as:

\begin{equation}
\frac{\dot{y}}{y} \equiv \frac{\dot{A}}{A} + \alpha \frac{\dot{k}}{k} \label{eq:shaikh}
\end{equation}
where $\frac{\dot{A}}{A} = [(1-\alpha)\dot{w}/w + \alpha \dot{r}/r]$.

Suppose, as it is roughly the case, that the shares of profits and wages are roughly constant with time.~\footnote{If wages increase, profits decrease. Thus, the economy slows down,  unemployment increases and, consequently, wages decrease. Thus, the economic system regulates itself around shares of profits and wages that  oscillate within bounds.} Then, $A$ is only a function of time and eq.~\ref{eq:shaikh} can be integrated to $y(t) \approx  A(t) k^\alpha$ which, by multiplying by $L(t)$, can be written as:

\begin{equation}
Y \approx A(t)\,K^\alpha \,L^{1-\alpha}    \label{eq:cobb-douglas-shaikh}
\end{equation}
which is \emph{not} a production function \cite{shaikh74}.

Equation~\ref{eq:cobb-douglas-shaikh} is not a production function in the sense that it does not express any technological capability. It simply arises out of definitional identities and an empirical regularity.

Thus, the Cobb-Douglas production function actually has no empirical support. Macroeconomic production functions of the form~\ref{eq:macro} may not exist at all. In spite of widely accepted simplifications, the intricacies of organizational relations may be unavoidable.

\section{Increasing Returns to Scale} \label{sec:increasing}

Economic theory assumes equilibrium, which is realized under the hypothesis of decreasing returns to scale. In the basic version the production function~\ref{eq:basic} is assumed, with $\partial f / \partial x_i >0 \; \forall i$ and $\partial^2 f / \partial x_i \partial x_j >0 \; \forall i,j: i\neq j$ but $\partial^2 f / \partial x_i^2 <0 \; \forall i$.

However, firms grow at differential rates and some of them become quite large. In order to explain this fact it is generally assumed that returns to scale may increase until size remains below a certain threshold, i.e. that $\forall i \: \exists \delta_i \geq 0: \; \forall x_i> \delta_i$ it is $\partial^2 f / \partial x_i^2 <0$ but  for at least one $i$, $\forall x_i< \delta_i$ it is $\partial^2 f / \partial x_i^2 >0$.

The distribution of the size of firms has be subjected to a number of empirical investigations. These have highlighted the following empirical regularities:

\begin{itemize}

\item The distribution of the size of firms is   skewed to the right \cite{ijiri-simon77}, deviates from a log-normal  \cite{stanley-buldyrev-havlin-mantegna-salinger-stanley95} and, according to the most recent studies, follows a power law \cite{axtell01} \cite{gaffeo-gallegati-palestrini03}.

\item The empirical investigations  carried out since the 1980s have ascertained that growth rates decrease with firm size \cite{hall87} \cite{dunne-hughes94}. In particular, the most recent studies fit the data with a Laplace distribution  \cite{amaral-buldyrev-havlin-leschhorn-maass-salinger-stanley-stanley97} \cite{amaral-gopikrishnan-plerou-stanley01} \cite{bottazzi-secchi03}.

\item The standard deviations of growth rates are known since a long time to decrease with size \cite{hymer-pashigian62} \cite{singh-whittington75}  \cite{hall87} \cite{dunne-hughes94}. By making use of more recent data it has been shown that the distribution of the standard deviations of growth rates follows a power law \cite{amaral-buldyrev-havlin-leschhorn-maass-salinger-stanley-stanley97} \cite{amaral-gopikrishnan-plerou-stanley01}.

\end{itemize}

 In principle, an appropriate choice of the distribution of production functions across industries and countries can explain any empirically observed distribution of the size of firms as well as the distributions of their growth rates and standard deviations. By means of an appropriate combination of increasing returns below a certain threshold and decreasing returns above it, any empirically observed distribution can be justified.

However, the observed empirical regularities span a period of several decades, in which  fundamental technological changes have taken place. Across decades, production functions have certainly changed. Thus, these regularities question the validity of received economic theory.

A right-skewed distribution can be produced by assuming that firms grow at a  rate independent of their size (Gibrat's law)  \cite{gibrat31}. However, the distribution of the size of firms would be log-normal, whereas some empirical analyses claim that it is not. Furthermore, growth rates and their standard deviations would be uniformly distributed, and they are not.

Dynamics of entry and exit  from the population of firms can be superimposed to Gibrat's law in order to account for the empirically observed regularities. Entry and exit dynamics  may result from internal restructurings of large organizations \cite{buldyrev-amaral-havlin-leschhorn-maass-salinger-stanley-stanley97} \cite{amaral-buldyrev-havlin-salinger-stanley98} \cite{plerou-amaral-gopikrishnan-meyer-stanley99} \cite{amaral-gopikrishnan-plerou-stanley01}, or from births and deaths of firms \cite{simon-bonini58} \cite{mccloughan95} \cite{levy-solomon96} \cite{solomon-levy96} \cite{blank-solomon00} \cite{richiardi04}.

Conventionally, size  is either measured in terms of output or number of employees. Let us assume that the number of employees (the amount of human capital) is proportional to output so production functions can take the form of eq.~\ref{eq:basic}. Under this condition, size is measured by output.

Gibrat's law translates into a stochastic multiplicative process  \cite{gibrat31}. The output  at time $t$  obtains from   the output at time $t-1$ via multiplication by a   $n \times n$ matrix of stochastic processes $\Lambda_{t-1}$:

\begin{equation} \label{eq:gibrat}
\mathbf{y}_{t} = \Lambda_{t-1} \: \mathbf{y}_{t-1}
\end{equation}

In order to avoid that the economy implodes, there must exist at least one eigenvalue $\lambda_i$ of $E\{\Lambda_{t-1} \}$ such that $|\lambda_i | \geq 1$. This implies that for at least one firms returns to scale are non-decreasing, a circumstance which is incompatible with general equilibrium theory. The idea is that the economy is inherently unstable and unpredictable, but that specific entry and exit dynamics cause the empirical regularities that have been observed in the distribution of the size of firms.

Economics does know a  mechanism that may provide regularizing entry and exit dynamics \cite{richardson60}.
When demand is increasing, firms increase their productive capacity at the fastest rate allowed by their profits and debts. In fact,  non-decreasing returns to scale imply that firms are playing a winner-takes-all game. They know that at a certain point demand will saturate so most of them will have excess productive capacity and high debts, but if they choose  not to grow, they will be doomed. By playing the game of growth, they have a chance to become the monopolist (most often, one of the few oligopolists that  share a saturated market). This is the exit dynamics. The entry dynamics becomes relevant after the loosers are out of the market and takes place on novel technologies.

An aggregate formalization of this mechanism was provided by Goodwin with his business cycle model where downswings begin when aggregate capital becomes larger than ``desired capital'', calculated as a fraction of income \cite{goodwin51} \cite{goodwin-punzo87}. The idea is that available demand calculated as a fraction of income constitutes a threshold for supply. If supply overcomes this threshold, some producers will go bankrupt. 

Let $y_t^i$ denote the output of firm $i$ at time $t$, with $i=1,2,\ldots n$. Let us suppose that there are $m$ markets, denoted by an index $j=1,2,\ldots m$, with $m \leq n$. Let $n(j)$ denote the number of firms that operate in market $j$.

Aggregate  income at time $t$ is $Y_t = \sum_{i=1}^{n} y_{t-1}^i$. This what consumers can purchase in all markets. Firms correctly predict that at time $t$ the size of market $j$ will be $c_t^j Y_{t-1}$, with $0 \leq c_t^j \leq 1$ for $\forall t,j$. Firms know that the market constraint is $\sum_{i=1}^{n(j)} y_t^i \leq c_t^j Y_{t-1}$. However, each firm $i$  operating in market $j$ makes its choices as if $y_t^i \leq c_t^j Y_{t-1}$.

To this exit dynamics, a random entry dynamics should be added. Thus, the rows of eq.~\ref{eq:gibrat} would change as follows:

\begin{enumerate}

\item For firms such that $\sum_{i=1}^{n(j)} y_t^i \leq c_t^j Y_{t-1}$, eq.~\ref{eq:gibrat} would remain unchanged.

\item For firms such that $\sum_{i=1}^{n(j)} y_t^i > c_t^j Y_{t-1}$, an exit dynamics should be devised that makes some of them disappear.

\item An entry dynamics should be added. Since this may create new firms as well as new sectors, in general $n_t \neq n_{t-1}$ and  $m_t \neq m_{t-1}$.

\end{enumerate}

Thus, in place of eq.~\ref{eq:gibrat} one would have:

\begin{equation} \label{eq:goodwin}
y_t^i = \left\{
\begin{array}{ll}
\Lambda_{t-1}^i \: \mathbf{y}_{t-1} & \mbox{ if  } \sum_{i=1}^{n(j)} y_t^i \leq c_t^j Y_{t-1} \\
\mathbf{A}_t \: \mathbf{y}_{t-1} & \mbox{ if  } \sum_{i=1}^{n(j)} y_t^i > c_t^j Y_{t-1} \\
\epsilon_{t-1}^i & \mbox{ for  } i > n_{t}
\end{array}
\right.
\end{equation}
where $\Lambda_{t-1}^i$, the $i-$th row of $\Lambda_{t-1}$, is such that $y_t^i \leq c_t^j Y_{t-1}$ and $\epsilon_{t-1}^i$ is another stochastic process.

Matrix $\mathbf{A}_t$  is an algorithm applied at time $t$ to the outcome of eq.~\ref{eq:gibrat} in order to cancel some firms until the requirement $\sum_{i=1}^{n(j)} y_t^i \leq c_t^j Y_{t-1}$  is satisfied. It may work by eliminating the smallest units first, or by reducing the output of all firms by a fixed proportion and eliminating those falling below zero output, or else.

Simulations of eq.~\ref{eq:goodwin} with thresholds provided by supply and demand have generated distributions of firm sizes very well, besides reproducing the observed stability of the size of industries \cite{richiardi04}. The idea of economies where increasing returns to scale are ubiquitous systemic constraints produce regular patterns is intellectally intriguing, technically challenging and empirically sensible.

\section{The authors put in context} \label{sec:context}

The previous sections expounded several critical topics surrounding production functions. Their place within economics, as well as  extent to which this paper provided original connections, can only be appreciated if the authors involved are framed within the history and the conventional content of this discipline.

Nicholas Georgescu-Roegen graduated in mathematics and physics before becoming an economist. Regrettably, his work on production functions is little known among economists. Since curricula in economics generally do not entail the concept of \emph{state variable}, even those economists who are acquainted with the flow-funds model are unable to view its dynamics as a general property of dynamical systems. Nobody seems to be aware that macroeconomic production functions are an instance of Georgescu-Roegen's flow-funds model.

Piero Sraffa, Joan Robinson and Pierangelo Garegnani were marxist~\footnote{More precisely, ``neo-Ricardians''.} economists who elaborated the theme of reswitching as a criticism to the neo-classical (bourgeois) theory of value. They had some resonance in the 1960s but were forgotten thereafter. Since they failed to recognize that the point they were making derives from a basic property of any function of two or more  variables, the debate on reswitching occupied tons of paper on the journals of that time.

The concept of an organizational production function is foreign to economics, as well as any connection between production functions and neural networks. Positive externalities are eventually included in production functions without any mention of structural issues.

Warren Hogan and Anwar Sheikh were applied economists. Their criticism of the Cobb-Douglas production function has been totally ignored by the profession, possibly because it is so destructive. Recently, Scott Moss made it known in the community of social scientists who make use of agent-based simulations.

The invariant properties of firms size distributions have been discovered by econophysicists and most often published on this journal.  These discoveries are  challenging because they question current economic theory. On the contrary, some paths that have been abandoned should be possibly resumed.

In particular, George Richardson is fundamental to understand economic dynamics without assuming that an equilibrium is necessarily there. His book was ignored when it was first published but it has been reprinted after more than twenty years.

Likewise, Richard Goodwin with his non-linear business cycle model is a prominent reference for non-equilibrium economic studies. Interestingly, Goodwin is also an early example of a physicist who chose economics as his research field.

\bibliographystyle{plain}
\bibliography{references}
\end{document}